\documentclass[english,aps,pra,reprint,noshowpacs,superscriptaddress,table]{revtex4-1}   
\usepackage[T1]{fontenc}	% should generally be included for better accented-word practices
\usepackage[latin9]{inputenc}	% should generally be included for better accent practices
\usepackage{geometry}		% for controlling page margins
\usepackage{graphicx}
\usepackage[above,below]{placeins}	% allows use of \FloatBarrier command to force section barriers
\usepackage{times}
\usepackage{hyperref}
\usepackage{array}
\hypersetup{colorlinks=true,urlcolor=blue,citecolor=blue,linkcolor=blue}   
\usepackage{multirow}
\usepackage{tabularx}

\usepackage[table]{xcolor}% http://ctan.org/pkg/xcolor
\definecolor{l0}{rgb}{1,1,1}
\definecolor{l1}{rgb}{0.95,1,0.95}
\definecolor{l2}{rgb}{0.9,1,0.9}
\definecolor{l3}{rgb}{0.85,1,0.85}
\definecolor{l4}{rgb}{0.8,1,0.8}
\definecolor{l5}{rgb}{0.75,1,0.75}
\definecolor{l6}{rgb}{0.7,1,0.7}
\definecolor{l7}{rgb}{0.65,1,0.65}
\definecolor{l8}{rgb}{0.6,1,0.6}
\definecolor{l9}{rgb}{0.55,1,0.55}
\definecolor{l10}{rgb}{0.5,1,0.5}

\usepackage{pgfplots}
\pgfplotsset{width=7cm,compat=1.14}
\usepackage{comment}

\usepackage{subcaption}

\usepackage{amsmath}

\begin{document}

\title{Confirming what we know: Understanding questionable research practices in intro physics labs}
\author{Martin M. Stein}
\author{Emily M. Smith}
\author{N. G. Holmes}
\affiliation{Laboratory of Atomic and Solid State Physics, Cornell University, 245 East Avenue, Ithaca, NY 14853}

%\keywords{}

\begin{abstract}
Many institutions are changing the focus of their introductory physics labs from verifying physics content towards teaching students about the skills and nature of science. As instruction shifts, so too will the ways students approach and behave in the labs. In this study, we evaluated students' lab notes from an early activity in an experimentation-focused lab course.
We found that about 30\% of student groups (out of 107 groups at three institutions) recorded questionable research practices in their lab notes, such as subjective interpretations of results or manipulating equipment and data. The large majority of these practices were associated with confirmatory goals, which we suspect stem from students' prior exposure to verification labs. We propose ways for experimentation-focused labs to better engage students in the responsible conduct of research and authentic scientific practice.
\end{abstract}

\maketitle

\section{Introduction}

\begin{table*}[!bth]
\caption{Different types of questionable research practices.}
\label{tab:new_biases}
\begin{tabularx}{\textwidth}{p{0.1\linewidth} l X} 
\hline\hline
Category & Questionable Research practices & Description\\
\hline

\multirow{4}{0.15\linewidth}{Subjective Interpretation} & \emph{Concerning results} &
The distinguishability of the datasets is described as a concern or issue. \\

& \emph{Emotional response to data} &
Statements refer to students liking or disliking the results. \\

&\multirow{2}{*}{\emph{Qualitative judgment of results}}  & 
The distinguishability of the data or quality of the methods are judged qualitatively (e.g. good, bad, too small, too large, helpful, or an improvement) based on the results. \\

\hline

\multirow{5}{0.15\linewidth}{Unjustified Interpretation}&\multirow{2}{*}{\emph{Claim of accuracy}} & 
The accuracy of data, the method, or an instrument used to take data, are judged based on the test statistic value. \\ 

&\multirow{2}{*}{\emph{Claim of systematic error}} & 
The distinguishability of the data sets is explained based on the presence or absence of systematic error, without describing the source.
\\ 

&\emph{Doubting statistics} & 
The validity of statistical tools, like the test statistic or standard deviation/error is questioned without justification.\\
\hline

\multirow{2}{*}{Purpose} & \multirow{2}{*}{\emph{(Dis-) prove model}} &
The purpose of the lab or intent of the group is explicitly or implicitly 
stated as to show that the model holds or breaks down.\\
\hline

Data manipulation & \multirow{2}{*}{\emph{Inflating uncertainty}} &
Statements demonstrate that students attempted to inflate their uncertainty, either through experimental decisions or manipulation of data.\\
\hline\hline
\end{tabularx}
\end{table*}

Introductory physics labs are often used to verify the physics content presented in a course.
These traditional, often highly structured, labs have been under heavy scrutiny. Studies have found that these labs do not provide measurable added value to learning the physics content~\cite{Holmes2017} and deteriorate students' perceptions of experimental physics~\cite{Wilcox2017DevelopingPhysics}. 

Rather than using labs to verify physics content, there are calls to shift the focus of labs to teach students about the nature of science and to develop students' scientific abilities. The American Association of Physics Teachers endorsed learning goals for labs that focus on a variety of experimentation skills and abilities~\cite{AAPT}. Encouragingly, studies investigating lab curricula centered around these goals, broadly referred to here as experimentation-focused labs, have found that students' abilities and engagement can develop over semester-long courses~\cite{Etkina2010DesignLaboratories,Holmes2015TeachingThinking}.
However, little is known about the struggles that students encounter in experimentation-focused labs. 

As labs shift away from verifying physics content, students may struggle to understand their role in the lab. In experimentation-focused labs, the focus is on the process, not the product, of the investigation. The intent is not for students to achieve a particular outcome; the intent is for students to work scientifically with data and models and to draw conclusions based on their evidence.

These instructional intentions may challenge students' beliefs about labs and the role of experiments in science. For example, many introductory students believe that the purpose of a lab is to supplement their learning of lecture content~\cite{Hu2018PERC,Hu2017}. Students also tend to believe that the purpose of experiments in physics labs is to confirm previously known results~\cite{Wilcox2017StudentsViews} and that experimental results should be evaluated on their agreement with theory or confirmation of previous results~\cite{Hu2018}.
We hypothesize that students' beliefs that labs are meant to confirm known results may prevent them from authentically and ethically engaging in the scientific process in experimentation-focused labs.

Previous research has found that some students interpret or manipulate data in ways that unjustifiably verify or confirm particular results---questionable research practices. For example, students were found to exhibit difficulties coordinating claims and evidence from complex data sets, such as making claims based on prior knowledge rather than data~\cite{Bogdan2014EffectsTables}. Students have also been found to inflate the values of experimental uncertainty to hide systematic errors that cause disagreement between data and theory~\cite{Holmes2014DoingExperiments}. 

In this paper, we identify that a large minority of students report questionable research practices during the first session of an experimentation-focused introductory physics lab course. We also find that the majority of students' questionable research practices were associated with a goal to confirm the provided model.
This study is part of a larger study that aims to evaluate how students transition to labs with no instructional intent to verify equations and where students have control over the outcome of their experiment.

\section{Methods}

Our participants were students enrolled in the first-semester course of a calculus-based physics sequence at three different institutions. The institutions included two research universities and one community college.
Data were collected from one semester at institutions A and B, but two separate semesters at institution C.

All four implementations  used the same activity during the first lab. This activity uses the Structured Quantitative Inquiry Labs (SQILabs) format~\cite{Holmes2015TeachingThinking}. In this activity, students are explicitly instructed to make comparisons between data (or data and a model), interpret those comparisons, and make decisions about how to follow-up on their investigation, with much emphasis on iterating to improve measurements.

Students worked in groups to conduct the investigations. Our data were the groups' written lab notes. They were instructed to record their process in a lab notebook and to treat the notebook like a journal, with extensive notes about what they were doing and why they were doing it. Students submitted the lab notebook by the end of the lab period for grading. At institution A and B, students had three hours to complete the lab. At institution C students had two hours to complete the lab. At institutions A and C students submitted one lab notebook per group, while students at Institution B submitted lab notes individually but only one student's notes were graded. Therefore, at all institutions, students received one grade for the whole group. At institutions A and B, students' notebooks were on paper, while at institution C, students used electronic lab notebooks.

Our data all stem from the first experiment of the lab course that asks students to test whether the period of a pendulum is dependent on its initial amplitude. Students were given the simplified formula for the period of a pendulum, $T=2\pi\sqrt{\ell/g}$, which predicts that the period is independent of the amplitude---valid under a small angle approximation. 
The approximation was not explicitly taught in the lab, though some students were aware of the equation's approximation.

During the lab, students were also introduced to a statistical test to distinguish the mean of two datasets within their uncertainties.
They were given an interpretation of the test statistic with three levels: a value below $1$ indicates statistically indistinguishable datasets, above $3$ means distinguishable, and in between, the result is unclear \cite{Holmes2015QuantitativeLab}.

Regardless of the result of the test statistic, students were encouraged to improve the quality of their measurements by iterating and extending the experiment. During the assigned lab time, many groups found statistically distinguishable datasets for the period at two different amplitudes (37\%, n=40), indicating they reached the experimental precision to measure the breakdown of the model.

\subsection{Development of the coding scheme}

We used emergent coding of the lab notes to identify questionable research practices. Table~\ref{tab:new_biases} summarizes the emergent codes and their categorization into more general questionable practices: subjective interpretations of data or methods, making unjustified interpretations of data or methods, explicitly stating an aim to prove or disprove the model, and manipulating data. There was most variability in the questionable ways groups coordinated claims and evidence, resulting in six different codes related to data interpretation.

Two raters coded 15\% of the notebooks. The selection included a greater proportion of questionable research practices than the entire dataset. Prior to discussion, raters agreed on average on 77\% of codes in each notebook. After simple wording changes and discussion, raters reached full agreement and a single rater coded the remaining lab notes.

Each coded questionable research practice was further divided as to whether or not it was oriented towards confirming the model. For example, one group stated an aim to make the periods more similar. This was coded with the questionable research practice \emph{(Dis-) prove model} and associated with confirming the model. Another group stated that their large value of the test statistic was a concern, which was coded with the questionable research practice \emph{Concerning results} and associated with confirming the model. Some groups recorded questionable research practices associated with both confirmatory and non-confirmatory goals, while other groups' practices had no clear goal.
For example, one group stated their value of the test statistic, which did not change between iterations, did not improve.
This was coded with the questionable research practice \emph{Qualitative judgment of results}, but not associated with confirmatory goals, because it was unclear what constituted an improvement.
Three groups indicated an aim to disconfirm the model. Because this aim involves a substantially different understanding of models and scientific exploration, these groups were not coded as aiming to confirm.

We further analyzed lab notes for groups' final result of the test statistic and the conclusions they drew from it. Our coding here matched the interpretation of the test statistic students were given: A test statistic smaller than 1 was coded as ``Test statistic supports model'', above 3 was coded as ``Test statistic contradicts model'' and in between as ``unclear''. The final conclusions in lab notes were coded as ``accept model'' if groups wrote the period of the pendulum is independent of the amplitude, ``reject model'' if they concluded there is a difference between the periods at different amplitudes, and ``unclear'' if it was explicitly written that the results were inconclusive or no definite conclusion was recorded.
We used these codes to identify what fraction of groups drew conclusions that agreed or disagreed with their data, and whether groups' confirmatory goals affected the conclusions they drew.

We performed an ordinal logistic regression to compare the groups with confirmatory questionable research practices and the groups with no coded questionable research practices. Deviations from the given interpretation of the test statistic were used as the ordinal response. Finding a test statistic that contradicts the model but accepting it in the conclusion was assigned the numerical value $-2$ and finding data that supports the model but rejecting the model in the conclusion was assigned $2$. Drawing a conclusion that followed the given interpretation of the test statistic was assigned $0$, starting from unclear data or arriving at an unclear conclusion was assigned $\pm 1$, depending on the orientation. 

\section{Results}

\begin{figure}[tb]
\includegraphics[width=\linewidth]{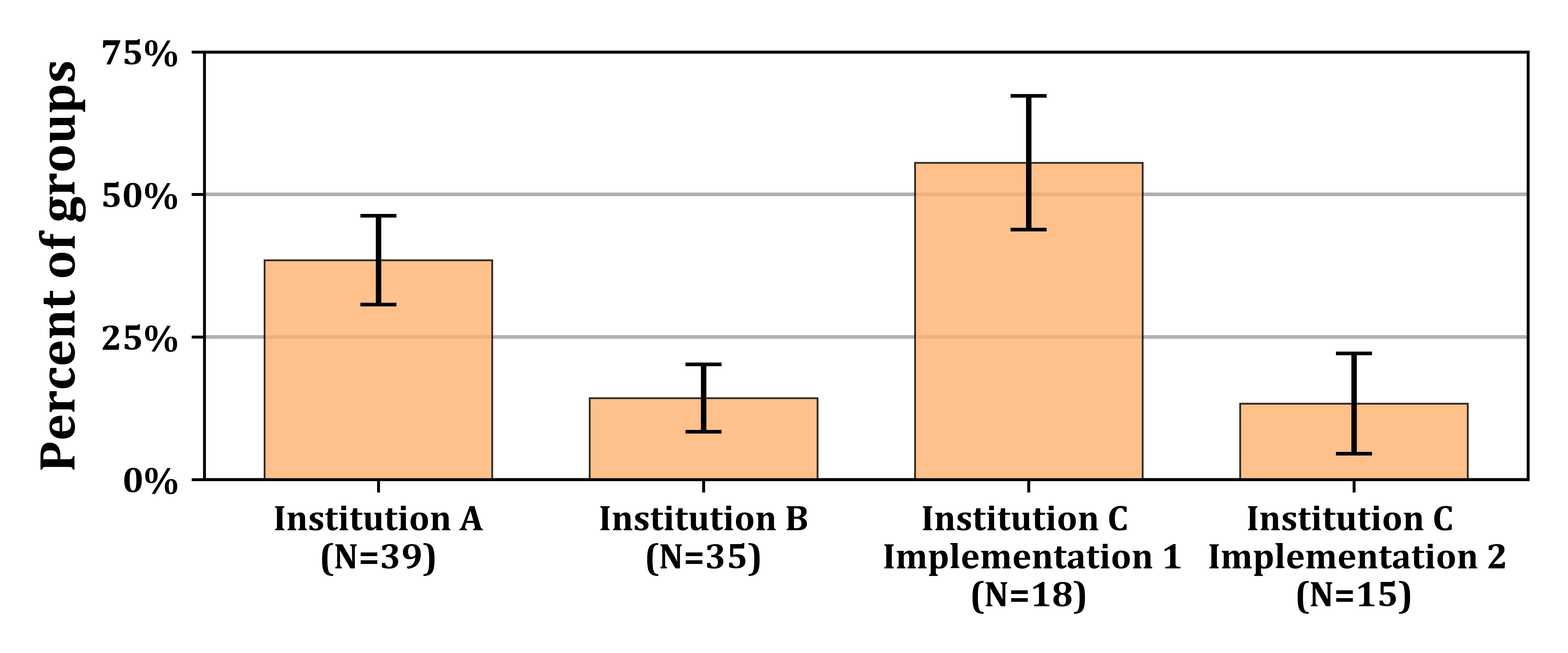}
\caption{Percent of groups that exhibited at least one questionable research practice across institutions.
Error bars represent standard errors on the proportions.
 \label{fig:BiasLevels}}
\end{figure}

Questionable research practices were found in 30\% of the groups' lab notes.
However, as shown in Fig.~\ref{fig:BiasLevels}, the fraction of lab notes indicating these practices varied significantly across implementations. We suspect, due to the significant differences in results between the two implementations at Institution C (Fisher's exact test: $p=0.013$), that variations in reported questionable practices are due to instructional decisions rather than student populations.

\begin{figure}[tb]
\includegraphics[width=\linewidth]{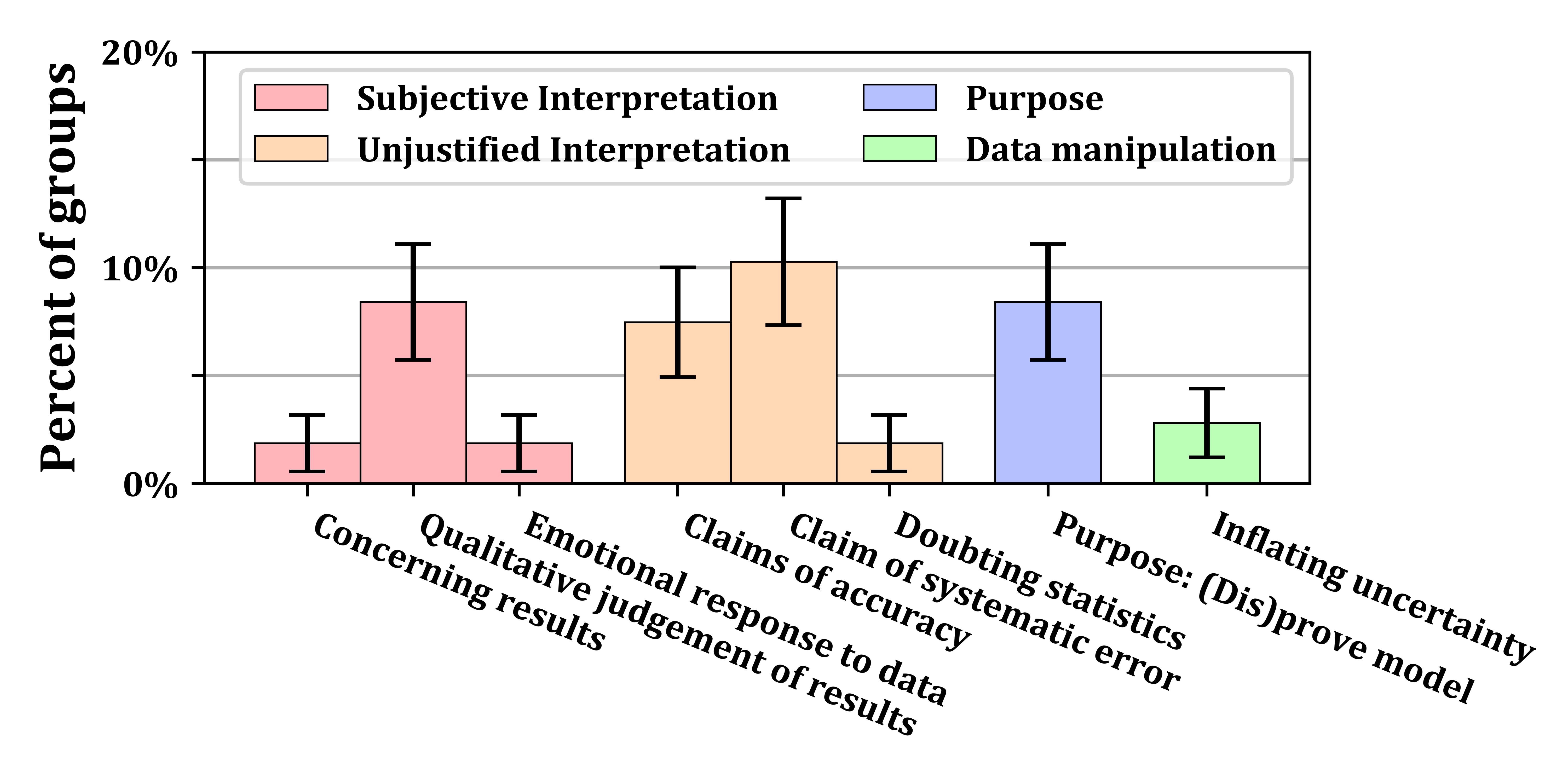}
\caption{Percent of groups that exhibited each questionable research practice. Error bars represent standard errors on the proportions.
}
\label{fig:BiasTypes}
\end{figure}

Some coded practices are noted much more frequently than others (Fig.~\ref{fig:BiasTypes}). Most commonly, groups noted practices that indicated subjective (12\% of all groups, n=13) or unjustified (20\% of all groups, n=21) interpretations of their data.
Most lab notes coded as unjustified interpretations made claims that their measurements were affected by a systematic error based on the test statistic.
For example, one group wrote, ``\emph{If we could reduce our error [...] our [test statistic] should have been much lower}.'' The next most common code in this category was from claims that the data or methods were accurate or inaccurate based on the agreement between data and the model. For example, one group wrote, ``\emph{This was a very crude experiment which resulted in a high [test statistic] which means our experiment was not very accurate.}''
Another group doubted that the statistical tools they used were representative of their data: ``\emph{Our [statistical] test is not representative of our data, this is because the standard error value is so large, giving us a lower [test statistic] value.}''

Many groups that recorded subjective interpretations of their results claimed the (dis-)agreement between data and the model was ``\emph{good}'', ``\emph{improved}'' or ``\emph{needs further help}'' (coded as \emph{Qualitative judgment of results}).
Others wrote the mismatch between their data and the model was a ``\emph{concern}''. The most surprising code was that some groups noted an emotional response to data that disagreed with the model. One group indicated they will change the experiment ``\emph{if there is any unsatisfaction}'' with their results, another group decided to increase their standard errors by an order of magnitude, because they ``\emph{liked the low [values of the test statistic]}'' and how indistinguishable the periods of the pendulum were with the larger uncertainties.

In fact, a few groups (3\%, n=3) actively manipulated their data by inflating experimental uncertainties to obtain results that agreed with the model.
This primarily involved groups designing follow-up experiments that deliberately increased their uncertainty to make the two datasets less distinguishable. Finally, about 8\% (n=9) of the groups stated that the purpose of the lab was to confirm or disconfirm the model or expressed their intent to do so.

Most (70\%) of the groups exhibiting questionable research practices, however, did so with a confirmatory goal.
An additional 20\% of the groups recording questionable research practices made statements that indicated they were trying to confirm the model, but also made statements that indicated the opposite or were unclear.
This implies that 90\% of the groups exhibiting questionable research practices conveyed confirmatory goals at some point.

\begin{figure}[h]
\begin{subfigure}{.5\linewidth}
  \centering
  \includegraphics[width=\linewidth]{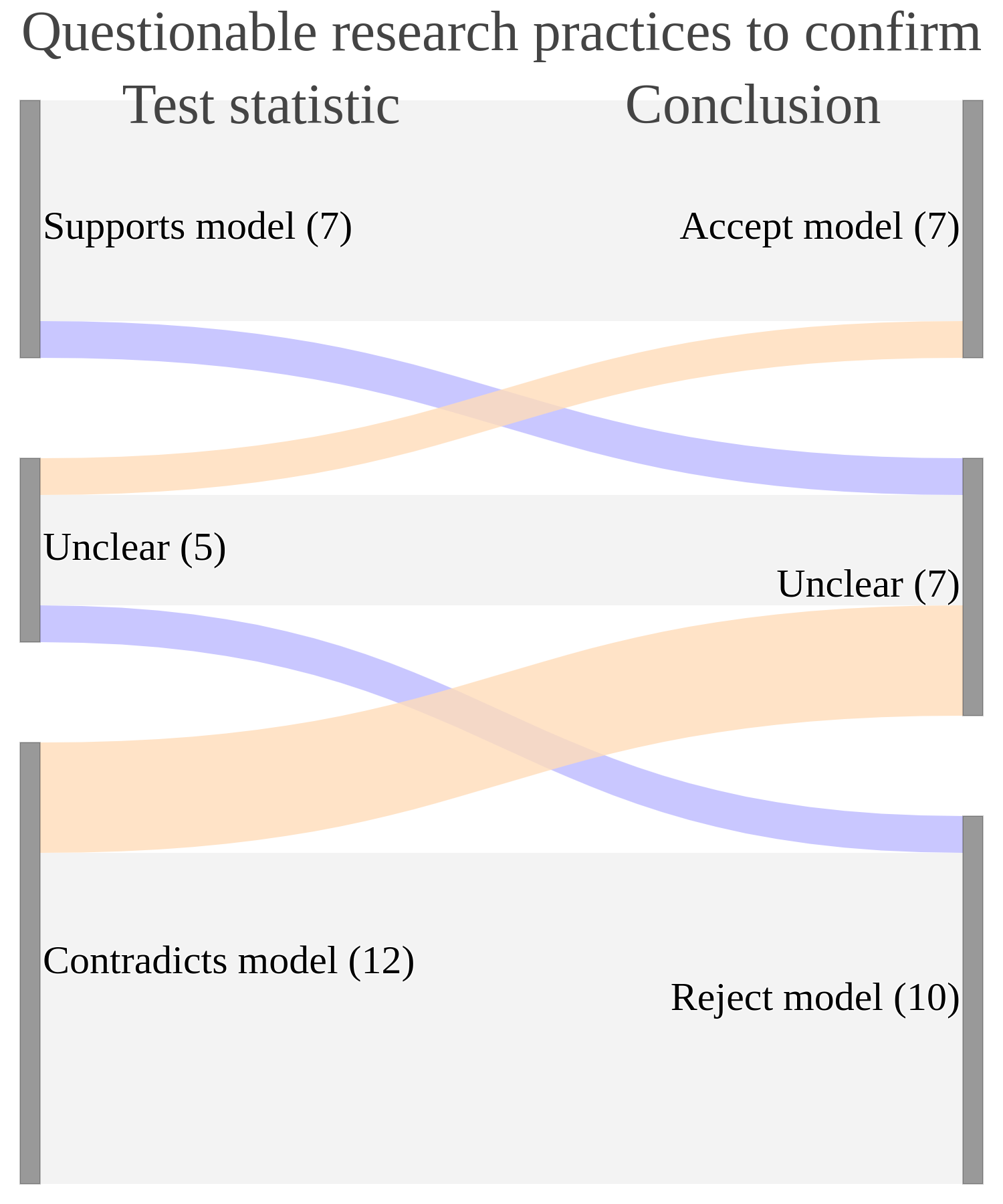}
  \label{fig:biased}
\end{subfigure}%
\begin{subfigure}{.5\linewidth}
  \centering
  \includegraphics[width=\linewidth]{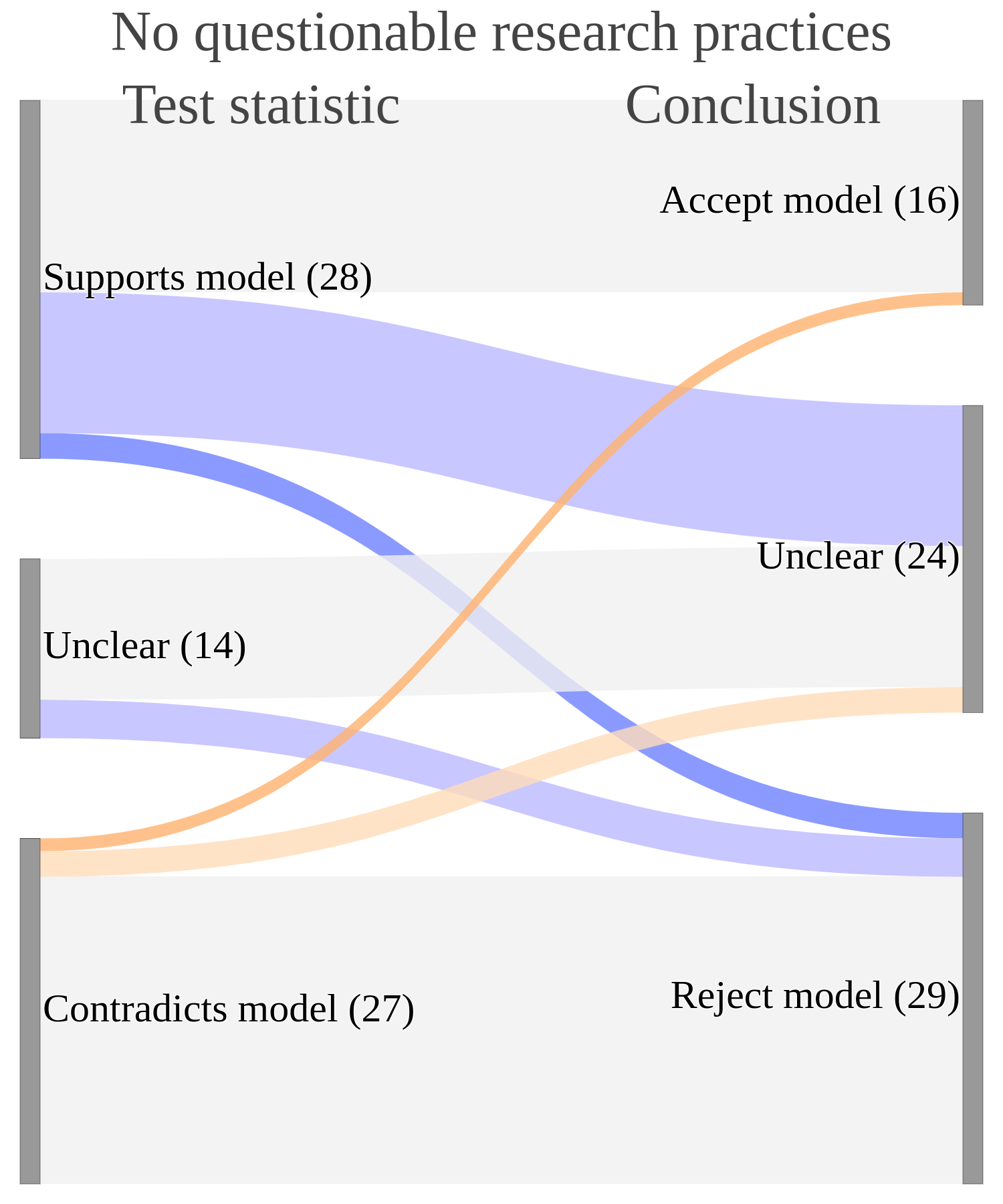}
  \label{fig:neutral}
\end{subfigure}
\caption{Results of test statistics (left nodes) and conclusions (right nodes) drawn by student groups using confirmatory questionable research practices vs. no questionable research practices. Orange lines correspond to conclusions that lean towards confirmation, blue lines towards disconfirmation, gray lines are directly supported by the test statistic obtained.}
\label{fig:sankey}
\end{figure}

In Fig.~\ref{fig:sankey} we compare students who exhibited confirmatory questionable research practices with those who did not exhibit questionable research practices. There were too few groups who exhibited non-confirmatory questionable research practices for meaningful comparison. Groups with confirmatory questionable research practices arrived at similar results as groups who did not exhibit questionable research practices, as measured by the values of their test statistic (Mann Whitney U-Test, $p=0.394$).
The two groups also arrived at similar conclusions at the end of the lab. However, as visualized in Fig.~\ref{fig:sankey}, groups exhibiting confirmatory questionable research practices less frequently questioned the model if they found data supporting it and less frequently rejected the model if they found data contradicting it. This can be seen in the ``upwards'' trend in the left diagram and the ``downwards'' trend in the right diagram.
The ordinal logistic regression found these differences statistically significant ($p= 0.022$). 

\section{Discussion}

In this study we found that students report questionable research practices in an experimentation-focused lab.
Most of these practices were associated with students' interpretation of data and served to confirm the investigated model.
Furthermore, students aiming to confirm the model less frequently questioned data that supported the model and less frequently rejected the model in light of contradicting data. 

Although we have only observed a correlation, we suspect that the intent to confirm the model was the cause of most questionable research practices. 
Research has found that the majority of introductory students believe that the validity of data should be evaluated based on its agreement with theory or the results from others \cite{Hu2018} and that the purpose of experiments is to confirm theory \cite{Hu2017}. We found that those confirmatory beliefs correlate with students' questionable research practices in the lab as many students motivated these practices with the intent to show agreement between data and the model.
In the future, we aim to investigate whether these confirmatory goals come from extensive previous experience with verification labs.

Our data only include those practices that students recorded in their lab notes. It is likely that more students engaged in additional practices but did not report them. The percentages we report, therefore can be understood as lower bounds on their occurrence in the labs.
The exhibited practices were similar across different institutions; however, we found their frequency varied significantly across different implementations. We attribute these variations to differences in instruction rather than student population, which could inform possible instructor interventions. Based on the results here, it is possible that clarifying the purpose of the lab to students, asking them to think critically about the limitations of simple models, or constantly strive to reduce uncertainty, could affect student practices.
However, it is unclear whether the differences in instruction actually changes the frequency of questionable research practices or just how often they are reported in lab notes. We plan to use video observation to investigate these and other questions in the future.
We present an analysis of video observations and interviews for how the presented lab activity can shift students framing of introductory physics labs away from model-verifying frames in \cite{SmithPERC}.

\acknowledgments{This work was partially supported by Cornell University's College of Arts \& Sciences Active Learning Initiative.}

\begin{comment}
% For short, simple bibliographies, manually formatting works:
% Remember that you'll need to run pdfLaTeX twice to get the references to show (first 
% pass will insert ?? in their places).

\end{comment}

% For a longer bibliography, delete the thebibliography block above, then comment in 
% these two lines to use a .bib file with BibTeX.
\bibliographystyle{apsrev}  	% supercedes the longbibliography option, so leave commented out if you want to display article titles
\bibliography{Mendeley.bib}  	% don't include the .bib suffix

\end{document}